\documentclass[smallextended]{svjour3}     
\smartqed  
\usepackage{graphicx}
\journalname{Astrophysics and space science}
\begin{document}

\title{Kelvin-Helmholtz instability in a weakly ionized layer}

\titlerunning{Kelvin-Helmholtz instability}        

\author{Mohsen Shadmehri \and Turlough P. Downes }

\authorrunning{Shadmehri \& Downes} 

\institute{M. Shadmehri \at
              School of Mathematical Sciences, Dublin City University, Glasnevin, Dublin 9, Ireland \\
              Department of Physics, School of Science, Ferdowsi University, Mashhad, Iran\\
              \email{mohsen.shadmehri@dcu.ie}           
           \and
 T. P. Downes \at
              School of Mathematical Sciences, Dublin City University, Glasnevin, Dublin 9, Ireland\\
              \email{turlough.downes@dcu.ie}
}

\date{Received: date / Accepted: date}

\maketitle

\begin{abstract}
We study the linear theory of Kelvin-Helmholtz instability in a
layer  of ions and neutrals with finite thickness. In the
short wavelength limit the thickness of the layer has a negligible
effect on the growing modes.  However, perturbations with wavelength
comparable to layer's thickness are significantly affected by the
thickness of the layer. We show that the thickness of the layer has
a stabilizing effect on the two dominant growing modes. Transition
between the modes not only depends on the magnetic strength, but
also on the thickness of the layer.

\end{abstract}

\section{Introduction}
\label{intro}

When two fluids are in relative motion on either side of a common
boundary, the Kelvin-Helmholtz (KH) instability can occur. This
instability, the simplest example of shear flow instability, is a
well-known phenomenon in fluid mechanics and astrophysics (e.g.\
Chandrasekhar 1961). For example, the interactions between the solar
wind and planetary magnetospheres are studied based on the KH
instability (e.g. Nagano 1979). The KH instability, because of the
velocity shear in the mass flow between coronal plume structure and the
interplume region of the sun, could be a potential source for some of
the Alfv\'enic fluctuations observed in the solar wind (Andries, Tirry \&
Goossens 2000). The structure and the helical wave motion in the ionized
cometary tails as studied by Ershkovich \& Mendis (1986), Ershkovich,
Prialnik \& Eviatar (1986) and  Ershkovich,  Flammer \& Mendis (1986) using
the KH theory. The KH instability theory has been successfully used to
interpret the structure of the pc-scale jet in the radio source 3C273
(Lobanov \& Zensus 2001). Birk et al.\ (2000) studied the role of KH
instabilities in superwinds of primeval galaxies. The effect of radiative
losses on the evolution of the KH instability in jets and outflows has been
studied by many authors, either in linear regime (e.g., Massaglia et al. 1992; Hardee \& Stone 1997) or by direct numerical simulations (e.g., Rossi et al.
1997; Downes \& Ray 1998). Recently, Michikoshi \& Inutsuka (2006) studied
the KH instability of the dust layer in protoplanetary discs to understand
the effect of relative motion between gas and dust.

However, in many astrophysical flows the single fluid approximation is
inappropriate. For example, the fact that the winds and the intergalactic
medium may be only partially ionized should be considered in treatment
of role of the KH instability in superwinds of primeval galaxies. We
may have a similar situation at the interface between a jet and its
cocoon if the KH instability is to be viewed as a possible mechanism
for entrainment. The problem becomes more complicated if we note that
the development of the KH instability is strongly influenced by the
magnetic field (e.g.\ Malagoli et al. 1996; Jones et al. 1997).
Therefore it seems a two-fluid treatment of KH instability, taking
account of the different motions of ions and neutrals and of the
magnetic field, rather than a pure one fluid MHD description is a more
appropriate approach.

By neglecting the thermal pressure forces of the neutrals or
perturbations in the neutral gas bulk velocities, the KH instability of
a system consisting of ions and neutrals with applications to partially
ionized cometary ionopauses, has been studied for both compressible
(Ershkovich \& Mendis 1986) and incompressible cases (Ershkovich,
Prialnik \& Eviatar 1986; Ershkovich, Flammer \& Mendis 1986). Chhajlani
\& Vyas (1991) incorporated the effects of finite resistivity and
rotation on the KH instability of two superposed fluids slipping past
each other. They showed that small rotation and the presence of neutral
particles destabilize the system. Birk et al.\ (2000) studied the KH
instability by taking into account the full incompressible dynamics of
both the neutral and the ionized gas components with applications to
multi-phase galactic outflow winds. In another similar study, Watson et
al.\ (2004) (hereafter WZHC) investigated the KH instability in the
linear, partially ionized regime to determine its possible effect on entrainment in massive bipolar outflows. They showed that for much of the relevant parameter space, neutral and ions are sufficiently decoupled that the neutrals are unstable while ions are held in place by the magnetic field. Birk \& Wiechen (2002) focused on unstable shear flows in partially ionized dense dusty plasma. They considered the dust and neutral gas components so that dust and neutral collisions is the dominant momentum transfer mechanism and dust component can interact with magnetic field lines, although dust charge fluctuations are negligible. They showed long wavelength modes can be stabilized by dust and neutral gas collisional momentum transfer. By doing multifluid numerical simulations, Wiechen (2006) studied KH modes in partially ionized, dusty plasmas for different masses and charges of dust.

Analytical studies of KH instability in weakly ionized medium have so
far considered one interface between two flows. For example, WHZC
studied incompressible KH instability in a two-fluid system (charged
and neutral) in order to understand entrainment in outflows from massive
stars. They did the calculations in  planar geometry with an interface at
$z=0$. However, it is not unreasonable to suppose that we may have a
flowing layer with finite thickness and two interfaces. In this paper,
we generalize the WHZC analysis by considering a finite thickness for a
weakly ionised flow. We will show that the growth time-scale of the
instability increases due to the thickness of the layer in particular
in the long wavelength limit. In the next section, we will present the basic equations and assumptions of the model. In the section 3, we will discuss  growing modes and their dependence on the thickness of the layer.

\section{General Formulation}
\label{sec:2}
We follow the analysis of Chandrasekhar (1961) but take account of the
different bulk velocities and densities of the neutral and ions, both
inside and outside the layer. These two components of our model are
coupled by collisional interactions. We assume incompressibility for
the analytical calculations and also that the convective term in the
induction equation dominates the resistive one. For doing linear analysis, the unperturbed properties of the system are important. We suppose that the streaming takes place in the $x-$direction with velocity $U(z)$,
\[U(z) = \left\{
\begin{array}{l l}
  +U & \quad \mbox{for $|z| \leq d$}\\
  -U & \quad \mbox{for $|z| > d$}\\ \end{array} \right. \]
($U$ is constant) and the neutral and ion components have the same
velocity in the unperturbed state. Note that $2d$ is the thickness
of the layer and $d\geq 0$. The magnetic field is assumed parallel
to the interface; that is, ${\bf B} = B_{x} {\bf e}_{x}$. Finally,
all unperturbed physical quantities are assumed constant in each
medium. The system under consideration consists of charged and
neutral fluid components coupled by collisional interactions. The
basic equations are

\begin{equation}
\nabla . {\bf v}_{n,i}=0,
\end{equation}
\begin{equation}
\rho_{n}(\frac{\partial{\bf v}_{n}}{\partial t}+ ({\bf v}_{n}.\nabla){\bf v}_{n})=-\nabla p_{n}- \gamma_{ni}\rho_{n}\rho_{i}({\bf v}_{n}-{\bf v}_{i})
\end{equation}
\begin{equation}
\rho_{i}(\frac{\partial{\bf v}_{i}}{\partial t}+ ({\bf v}_{i}.\nabla){\bf v}_{i})=-\nabla p_{i}+\frac{1}{4\pi}(\nabla\times{\bf B})\times {\bf B}- \gamma_{ni}\rho_{i}\rho_{n}({\bf v}_{i}-{\bf v}_{n})
\end{equation}
\begin{equation}
\frac{\partial {\bf B}}{\partial t}=\nabla\times ({\bf v}_{i}\times{\bf B}).
\end{equation}
Also note that $\nabla.{\bf B}=0$. In the above equations, $\gamma_{ni}$ is the collision rate coefficients per unit mass so that $\nu_{ni}=\gamma_{ni}\rho_{i}$ is the neutral-ion frequency. The collision frequency determine the scale of the coupling between the different components and the coupling between each component and the magnetic field.

Now, we perturb the physical variables as $\chi (z,x,t) = \chi'(z) \exp[i(\omega t + k_{x}x)]$, and so
\begin{equation}
-k_{x}\rho_{ n} u_{n}' + i \rho_{n}\frac{dw_{n}'}{dz}=0,
\end{equation}
\begin{equation}
\phi\rho_{n}u_{n}'=-k_{x}p_{n}'+i\gamma_{ni}\rho_{n}\rho_{i}(u_{n}'-u_{i}'),
\end{equation}
\begin{equation}
\phi\rho_{n}w_{n}'=i\frac{dp_{n}'}{dz}+i\gamma_{ni}\rho_{n}\rho_{i}(w_{n}'-w_{i}'),
\end{equation}
where $\phi=\omega + k_{x}U$, and $u_{n}'$ and $w_{n}'$ are the $x$ and the $z$ components of the perturbed velocity of the neutrals.  Also, the equations of the ions are
\begin{equation}
-k_{x}\rho_{ i} u_{i}' + i \rho_{i}\frac{dw_{i}'}{dz}=0,
\end{equation}
\begin{equation}
\phi\rho_{i}u_{i}'=-k_{x}p_{i}'+i\gamma_{ni}\rho_{n}\rho_{i}(u_{i}'-u_{n}'),
\end{equation}
\begin{equation}
\phi\rho_{i}w_{i}'=i\frac{dp_{i}'}{dz}-\frac{B_{x}^{2}}{4\pi\phi}\frac{d^{2}w_{i}'}{dz^{2}}+\frac{k_{x}^{2}B_{x}^{2}}{4\pi}\frac{w_{i}'}{\phi}
+i\gamma_{ni}\rho_{n}\rho_{i}(w_{i}'-w_{n}'),
\end{equation}
where $u_{i}'$ and $w_{i}'$ are the $x$ and the $z$ components of the perturbed velocity of the ions. We can reduce the above differential equations to a set of two
differential equations for $w_{i}'$ and  $w_{n}'$  as follows (WZHC)
\begin{equation}\label{eq:main2}
(\rho_{i}\phi - \frac{B^{2}k_{x}^2}{4\pi\phi}){\mathcal D}w_{i}'=i\gamma_{ni}\rho_{n}\rho_{i}{\mathcal{D}}(w_{i}'-w_{n}'),
\end{equation}
\begin{equation}\label{eq:main1}
\rho_{n}\phi {\mathcal{D}}w_{n}'=i\gamma_{ni}\rho_{n}\rho_{i}{\mathcal{D}}(w_{n}'-w_{i}'),
\end{equation}
where $\mathcal{D}\equiv \frac{d^{2}}{dz^{2}}-k_{x}^{2}$. The behavior
of the flow both inside and outside the layer is determined by the
general solutions of the linear differential equations (\ref{eq:main1})
and (\ref{eq:main2}). One can easily show that the general solutions of these equations are linear combinations of $\exp(+k_{x}z)$ and $\exp(-k_{x}z)$. We impose the condition that the perturbed quantities do not diverge at $z=+\infty$ and $z=-\infty$. Thus, the general solutions are
\begin{equation}w_{i}'(z) = \left\{
\begin{array}{l l}
  C_1 e^{+k_{x}z} +  C_{2} e^{-k_x z}& \quad \mbox{for $0 \leq z \leq d$}\\
  C_3 e^{-k_{x}z} & \quad \mbox{for $d \leq z$}\\ \end{array} \right.
\end{equation}

\begin{equation}w_{n}'(z) = \left\{
\begin{array}{l l}
  C_4 e^{+k_{x}z} +  C_{5} e^{-k_x z}& \quad \mbox{for $0 \leq z \leq d$}\\
  C_6 e^{-k_{x}z} & \quad \mbox{for $d \leq z$}\\ \end{array} \right.
\end{equation}
where $C_1$, $C_2$, $C_3$, $C_4$, $C_5$ and $C_6$ are constants which
are determined by the boundary conditions. We note that, because of
symmetry of the unperturbed state, there are solutions where $w_{i}(z)
    = - w_{i} (-z)$ and $w_{n}(z) = - w_{n} (-z)$ (odd solutions) and solutions where $w_{i}(z) = w_{i} (-z)$ and $w_{n}(z)
    = w_{n} (z)$  (even solutions). Thus, we can write the boundary condition at $z=0$ as
\begin{equation}
C_{1} + S C_{2} = 0,\label{eq:bound1}
\end{equation}
\begin{equation}
C_{4} + S C_{5}=0,\label{eq:bound2}
\end{equation}
where $S = +1$ and $S= -1$ are corresponding to even and odd solutions, respectively. Thus, considering symmetry properties of the solutions (i.e. equations (\ref{eq:bound1}) and (\ref{eq:bound2})) our solutions are also valid for $z<0$, if we transform $z\rightarrow -z$. So, the solutions correctly vanish, as $z\rightarrow -\infty$.

Another boundary condition is the continuity of the ion and neutral displacements across the interfaces. So, the boundary conditions at $z=d$ are
\begin{equation}
\frac{C_1 e^{k_x d} + C_2 e^{-k_x d}}{\omega + k_x U}=\frac{C_3 e^{-k_x d}}{\omega - k_x U},\label{eq:bound3}
\end{equation}
\begin{equation}
\frac{C_4 e^{k_x d} + C_5 e^{-k_x d}}{\omega + k_x U}=\frac{C_6 e^{-k_x d}}{\omega - k_x U}.\label{eq:bound4}
\end{equation}
Having equations (\ref{eq:bound1}), (\ref{eq:bound2}), (\ref{eq:bound3}) and (\ref{eq:bound4}), we can rewrite $w_i$ and $w_n$ as
\begin{equation}w_{i}'(z) = \left\{
\begin{array}{l l}
  C_2 (e^{-k_x z} - S e^{k_x z})& \quad \mbox{for $0 \leq z \leq d$}\\
  C_2 \frac{\omega - k_x U}{\omega + k_x U} (1-S e^{2k_{x}z}) & \quad \mbox{for $d \leq z$}\\ \end{array} \right.\label{eq:C2}
\end{equation}

\begin{equation}w_{n}'(z) = \left\{
\begin{array}{l l}
  C_5 (e^{-k_x z} - S e^{k_x z})& \quad \mbox{for $0 \leq z \leq d$}\\
  C_5 \frac{\omega - k_x U}{\omega + k_x U} (1-S e^{2k_{x}z})& \quad\mbox{for $d \leq z$}\\ \end{array} \right.\label{eq:C5}
\end{equation}

Clearly, the differential equations (\ref{eq:main1}) and (\ref{eq:main2}) are not valid at the interfaces $z=d$ (or $z=-d$), where velocities of the species are not analytic. However, we can integrate each equation over an infinitesimal region around the interface $z=d$ and use standard Gaussian pillbox arguments. Then we have
\begin{equation}
\triangle \{\rho_{n}\phi \frac{dw_{n}'}{dz}\}=i\triangle \{\gamma_{ni}\rho_{n}\rho_{i} \frac{d}{dz}(w_{n}'-w_{i}') \},
\end{equation}
\begin{equation}
\triangle \{\rho_{i}\phi \frac{dw_{i}'}{dz}\}=i\triangle \{\gamma_{ni}\rho_{n}\rho_{i} \frac{d}{dz}(w_{i}'-w_{n}') \}+\frac{B_{x}^{2}k_{x}^{2}}{4\pi}\triangle\{\frac{1}{\phi}\frac{dw_{i}'}{dz}\},
\end{equation}
By substituting equations (\ref{eq:C2}) and (\ref{eq:C5}) into the above equations and after lengthy (but straightforward) mathematical manipulations, we obtain
\begin{equation}
[\xi^2 + m\xi - (1-\lambda^2)h^2 + a^2] C_2 - m\xi C_5 =0,\label{eq:C2main}
\end{equation}
\begin{equation}
-\xi C_2 + [\xi^2 + \xi - (1-\lambda^2)h^2] C_5 =0,\label{eq:C5main}
\end{equation}
where
\begin{equation}
m=\frac{\rho_n}{\rho_{i}}, h = \frac{Uk_x}{\nu}, a = \frac{k v_{Ai}}{\nu}, \frac{\omega}{\nu}-\lambda h = -i\xi,
\end{equation}
and $\lambda = S^{-1}\exp (-2kd)$ and the ion Alfv\'en velocity $v_{Ai}$ is defined as $B/\sqrt{4\pi \rho_i}$. By the condition that the algebraic equations (\ref{eq:C2main}) and (\ref{eq:C5main}) have a nontrivial solution, we obtain the dispersion relation
\begin{displaymath}
\xi^4 + (m+1) \xi^3 - [2(1-\lambda^2)h^2-a^2]\xi^2
\end{displaymath}
\begin{equation}
-[(m+1)(1-\lambda^2)h^2-a^2]\xi + (1-\lambda^2)[(1-\lambda^2)h^2-a^2]h^2=0.\label{eq:mainm}
\end{equation}
This algebraic equation (\ref{eq:mainm}) is the basis of our stability analysis.
As we see, the effect of the thickness of the layer appears by term $\lambda^2$ which depends on the thickness of the layer. For a system with one interface at $z=0$, WZHC obtained this algebraic equation
\begin{displaymath}
\xi^4 + (m+1) \xi^3 - (2h^2-a^2)\xi^2-[(m+1)h^2-a^2]\xi + (h^2-a^2)h^2=0.
\end{displaymath}
We note that our algebraic equation (\ref{eq:mainm}) reduces to the above equation if we set $\lambda = 0$.

\begin{figure}
\includegraphics[width=12cm]{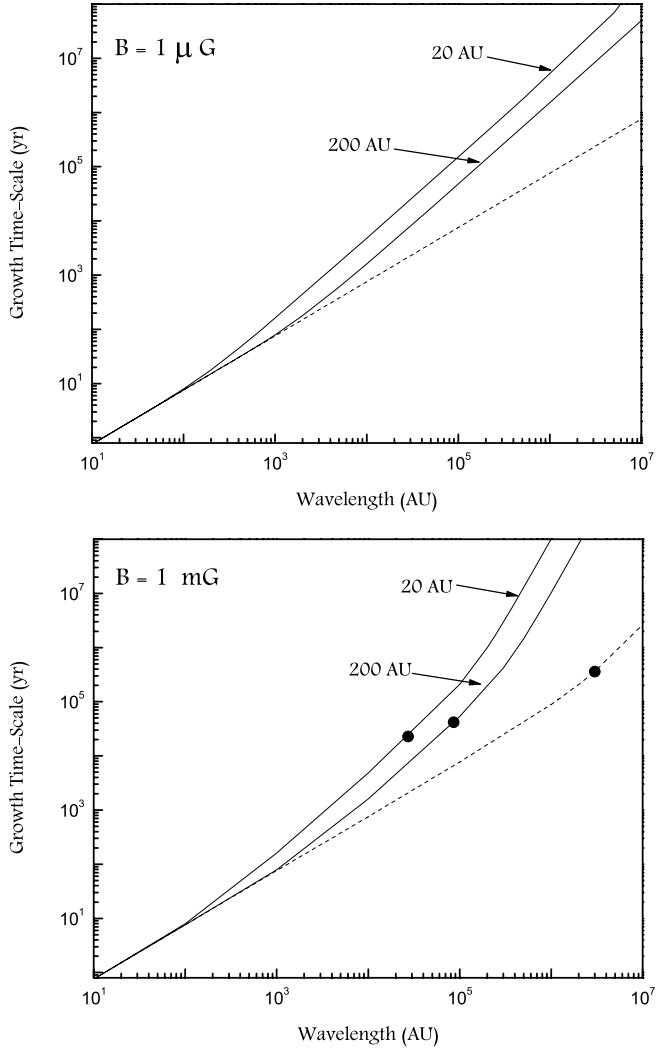}
\caption{Growth rate of the instability vs. wavelength of the
perturbations for two cases $B=1$ $\mu$G ({\it top}) and $B = 1$ mG
({\it bottom}). In both cases, it is assumed $U=10$ Km s$^{-1}$. The
dashed curves correspond to the WZHC analysis and the other
curves are labeled by the thickness of layer. Black circles in the
bottom plot denote the wavelength at which the growing mode changes from
$(1-\lambda^2)^{1/2}h$ to $(1-\lambda^2)h^2$.} \label{fig:1}
\end{figure}

\section{Analysis}
\label{sec:3}
Roots of equation (\ref{eq:mainm}) with positive real parts correspond to
growing unstable modes. This equation is applicable for various
weakly ionized systems. However, in this work the input parameters are chosen
to be appropriate for molecular outflows. This makes for easier comparison
with the analysis of WZHC. In our illustrative examples, the relative
velocity $2U$ is 20 Km s$^{-1}$, and the ratio, $m$, of neutral to
ionized density is assumed to be around $10^6$. The collision frequency is
$\nu=1.5\times 10^{-13}$ s$^{-1}$ (Draine et al.\ 1983). The magnetic
field strength is taken to be either $1$ $\mu$G (weak-field case) or
$1$ mG (strong-field case). With these parameters the ratio of the
Alfven velocity to the relative velocity is 4400 and 4.4 in the strong and weak magnetic field cases, respectively.  Having these input parameters, we can solve the algebraic equation (\ref{eq:mainm}) numerically to find the
dispersion relation and the fastest growing modes.

Figure \ref{fig:1} shows plots of the dispersion relation for the
above input parameters. When $\lambda \rightarrow 0$, the dispersion
relation tends to WZHC analysis (dashed curves in Figures \ref{fig:1}),
as expected. Each curve is also labeled by the thickness of
the layer, i.e. $2d$. It can be seen from Figure \ref{fig:1} that the effect
of the thickness of the layer is not significant for perturbations with
short wavelength. However, as the
wavelength of the perturbations increases, the thickness of the layer has a
stabilizing effect on the growing modes and the dispersion relation deviates
from the analysis of WZHC both in the weakly and strongly magnetised
cases. Considering the thickness of the layer implies longer growth
time-scale for the perturbations, in particular in the long wavelength
limit. Moreover, as one would expect, as the thickness of the layer
decreases the deviation from the WZHC analysis occurs at shorter
wavelength, and also the growth time-scale becomes longer.

Interestingly, the roots of equation (\ref{eq:mainm}) which correspond
to growing modes can be described by approximate analytical
solutions. There are three approximate positive roots for
this equation depending on its coefficients:

\noindent (i) $(1-\lambda^2)^{1/2}h$, when $(1-\lambda^2)^{1/2}h >
1$;

\noindent (ii) $(1-\lambda^2)h^2$, when $(1-\lambda^2)^{1/2}h < 1$
and $(1-\lambda^2)^{1/2}  < (a/h)m^{-1/2}$;

\noindent (iii) $(1-\lambda^2)^{1/2}h
(1-(a/h)^2(1-\lambda^2)^{-1}m^{-1})^{1/2}$ and $(1-\lambda^2)^{1/2}
> (a/h)m^{-1/2}$.

The agreement between these  approximate roots and numerical roots
of the equation (\ref{eq:mainm}) is excellent. Thus, as long as
$(1-\lambda^2)^{1/2}h > 1$ the only growing mode is the first
growing mode, irrespective of the strength of the magnetic field
(see top plot of Figure \ref{fig:1}). Once this inequality is violated,
we may have the second growing mode, depending on whether
$(1-\lambda^2)^{1/2}  < (a/h)m^{-1/2}$. For example, for $B=1$ mG we have
$(a/h)m^{-1/2}=4.5 > 1 \geq (1-\lambda^2)^{1/2}$ and so the transition from
the first to the second growing modes occurs when $(1-\lambda^2)^{1/2}h = 1$.
We can denote this transition wavelength by $\lambda_{t}$ and is
shown by black circles in Figure \ref{fig:1} (bottom plot). But in
the weakly magnetic case where $B=1$ $\mu$G, we have
$(a/h)m^{-1/2}=4.5\times 10^{-3} < (1-\lambda^2)^{1/2}$. So, for
wavelengths greater than $\lambda_{t}$, the third growing mode
appears, although the second term inside the parenthesis of this
root is much smaller than unity and the third mode actually tends
toward first mode (top plot). Considering our ranges of the input
parameters, the third mode is very close to the first mode.

\begin{figure}
  \includegraphics[width=12cm]{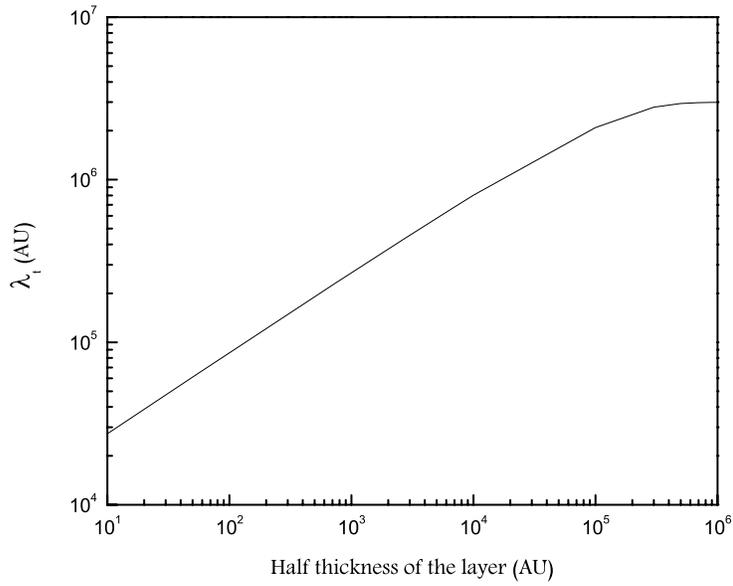}
\caption{The transition wavelength $\lambda_{t}$ vs. the half
thickness $d$ of the layer.} \label{fig:2}
\end{figure}

The first mode corresponds to the usual hydrodynamic instability, but the
second mode appears in our weakly ionized case as a new mode (WZHC). The transition between these two
modes depends on the thickness of the layer.  Figure \ref{fig:2} shows the
transition wavelength $\lambda_{t}$ versus the the half thickness of the
layer. For thin layers, the transition wavelength is smaller than
the thick layers.  In fact, as the thickness of the layer increases, this
transition wavelength increase and tends to an asymptotic value for the very
thick layers. WZHC analysis corresponds to this asymptotic regime.

\section{Conclusions}
\label{sec:4}
In this study we have extended the analysis of WZHC by considering a
layer of ions and neutrals with finite thickness interacting with an
ambient medium in the incompressible limit. However, this model can not
be applied directly for systems, such as outflows or jets because of
our simplifying assumptions. We can,
however, gain insight into the possible effects of shear between a moving
layer of ions and neutrals and the surrounding medium.

To our knowledge, there are a few numerical simulations of KH
instability of a multifluid system (Birk et al. 2000; Birk \&
Wiechen 2002; Wiechen 2006). While Birk et al. (2000) showed that KH
modes can operate fast enough to amplify the magnetic field
strengths in superwinds of primeval galaxies within the timescale of
the outflow, Birk \& Wiechen (2002) and Wiechen (2006) studied KH
instability in a dusty magnetized medium. It is difficult to make a
direct comparison between our results and these numerical
simulations because of different basic assumptions. We considered an
incompressible two-fluid layer, but all these numerical simulations
are for compressible systems. Moreover, the effect of charged dust
particles are considered in Birk \& Wiechen (2002) and Wiechen
(2006), though we have neglected charged grains in our analysis.

We have shown that the growth time-scale of KH instability in a layer
of ions and neutrals increases when compared to a system with one
interface, this effect being more evident in the long wavelength limit.
There are two dominant growing modes of KH instability in such a
system, depending on the magnetic strength and the thickness of the
layer. As the thickness of the layer decreases, the transition between
these two unstable modes occurs at shorter wavelengths. Given these
results, it remains to future work to extend this analysis to compressible case and make
 direct comparisons with jets or outflows with finite thickness.

\begin{acknowledgements}
The research of M. S. was funded under the Programme for Research in
Third Level Institutions (PRTLI) administered by the Irish Higher
Education Authority under the National Development Plan and with partial
support from the European Regional Development Fund.
\end{acknowledgements}



\end{document}